\documentclass[12pt,preprint]{aastex}

\usepackage[usenames]{color}

\usepackage{graphicx}
\usepackage{epstopdf}
\usepackage{url}

\newcommand{\be}{\begin{equation}}
\newcommand{\ee}{\end{equation}}
\newcommand{\ba}{\begin{eqnarray}}
\newcommand{\ea}{\end{eqnarray}}
\newcommand{\bc}{\begin{center}}
\newcommand{\ec}{\end{center}}
\newcommand{\lsi}{LS~I~$+$61$^{\circ}$303}

\newcommand{\swift} {\emph{Swift}}
\newcommand{\rxte}  {\emph{RXTE}}

\def \cm2{cm$^{-2}$\,}

\def\ergscm2 {erg\,s$^{-1}$cm$^{-2}$}

\begin{document}

\shorttitle{\textsc{The super-orbital modulation of \lsi\ in X-rays}}
\shortauthors{\textsc{Li et al.}}

\title{\textsc{Unveiling the super-orbital modulation of \lsi\ in X-rays}}

\author{Jian Li\altaffilmark{1},
Diego F. Torres\altaffilmark{2,3}, Shu Zhang\altaffilmark{1},
Daniela Hadasch\altaffilmark{2}, Nanda Rea\altaffilmark{2}, G. Andrea Caliandro\altaffilmark{2},
Yupeng Chen\altaffilmark{1}, Jianmin Wang\altaffilmark{1} }

\altaffiltext{1}{Laboratory for Particle Astrophysics, Institute of High
Energy Physics, Beijing 100049, China. Email: jianli@ihep.ac.cn}
\altaffiltext{2}{Institut de Ci\`encies de l'Espai (IEEC-CSIC),
              Campus UAB,  Torre C5, 2a planta,
              08193 Barcelona, Spain}
\altaffiltext{3}{Instituci\'o Catalana de Recerca i Estudis Avan\c{c}ats (ICREA).}

\begin{abstract}

From the longest monitoring of \lsi\ done to date by the  \textit{Rossi X-ray Timing Explorer} (\textit{RXTE})  we  found evidence for the long-sought,
years-long modulation in the X-ray emission of the source.
The time evolution of the modulated fraction in the
orbital lightcurves can be well fitted with a sinusoidal function having a super-orbital period of 1667 days, the same as the one reported in non-contemporaneous radio measurements.
However, we have found a  281.8 $\pm$ 44.6 days shift between the super-orbital variability found at radio frequencies extrapolated to the observation time of our campaign and what we found in the super-orbital modulation of the modulated fraction of our X-ray data.
We also find a super-orbital modulation in the maximum count rate of the orbital lightcurves, compatible with the former results, including the
shift.

\end{abstract}

\keywords{X-rays: binaries, X-rays: individual (\lsi)}

\section{Introduction}

\lsi\ is one of a handful high-mass X-ray binaries that have been detected at all frequencies, including TeV and GeV energies.
Its nature is still under debate, with rotationally-powered pulsar-composed systems (see Maraschi \& Treves 1981, Dubus 2006) and microquasar jets (see Bosch-Ramon \& Khangulyan 2009 for a review) being discussed. Recently, evidence favoring \lsi\
as the source of a very short X-ray burst led to the analysis of a third alternative, in which \lsi\ is a magnetar binary (see Torres  et al. 2011, also Bednarek 2009, and Dubus 2010). Long-term monitoring of the source, ideally across all wavelengths, is a key ingredient to disentangle differences in behavior which could point to the underlying source nature.

Here, we report on the analysis of  \rxte-Proportional Counter Array (PCA) monitoring observations of the $\gamma$-ray binary system \lsi.
The dataset we consider covers the period between
2007 September and 2011 September (2007-08-28 -- 2011-09-15), and it is the longest monitoring campaign done for  this source. Smaller datasets included in this campaign have been analyzed and results have been presented in our previous papers I and II (Torres et al. 2010, and Li et al. 2011, respectively). In them, we focused on establishing the orbit-to-orbit X-ray variability
and on studying the spectral properties and the flares. The current data enhances in more than one
additional year the reported coverage on the source, and for the first time, it is sufficient to consider
the possible super-orbital modulation of the X-ray emission from \lsi.

\section{Observations, data analysis, and results}

Our current analysis includes 473 \rxte-PCA pointed observations identified by proposal numbers 93100, 93101, 93102, 93017, 94102, 95102 and 96102.
The analysis of PCA data was performed using HEASoft
6.9. We filtered the data using the standard \rxte-PCA criteria.
Only PCU2 (in the 0-4 numbering scheme) has
been used for the analysis because it was the only Proportional
Counter Unit (PCU) that was 100\% on during the observations.
We select time intervals where the source elevation is $>$10$^{\circ}$ and
the pointing offset is $<0.02^{\circ}$. The background file used in the
analysis of PCA data is the most recent available one from the
HEASARC Web site for faint sources, and detector breakdown
events have been removed.\footnote{The background file is
{\tt pca\_bkgd\_cmfaintl7\_eMv20051128.mdl}.
The data have been barycentered using the {\tt
FTOOLS} routine {\tt faxbary} using the JPL DE405 solar system ephemeris. }
Our flux and count rate values are given for an energy range of 3--30 keV.

At first, we consider the complete
X-ray lightcurve of our campaign, and fitted it with a constant to obtain an average flux, resulting in (1.616$\pm$0.006) counts s$^{-1}$.
In order to remove
the influence of the several ks-long flares detected from the source,
we cut all observations that presented a count rate larger than three times this average.
The remaining on-source time amounts to 684.3 ks (460 observations), and it is uniformly distributed in the system's orbital (between 60 and 80 ks per each 0.1 of orbital phase bin)
as well as in the system's super-orbital phase (around 60 ks per each 0.1 of super-orbital phase bin, except at super-orbital phase 0.8, where an intensive campaign increased the exposure to 120 ks) as defined by the radio observations (Gregory 2002).

\begin{figure*}[t]
\centering
  \includegraphics[angle=0, scale=0.4] {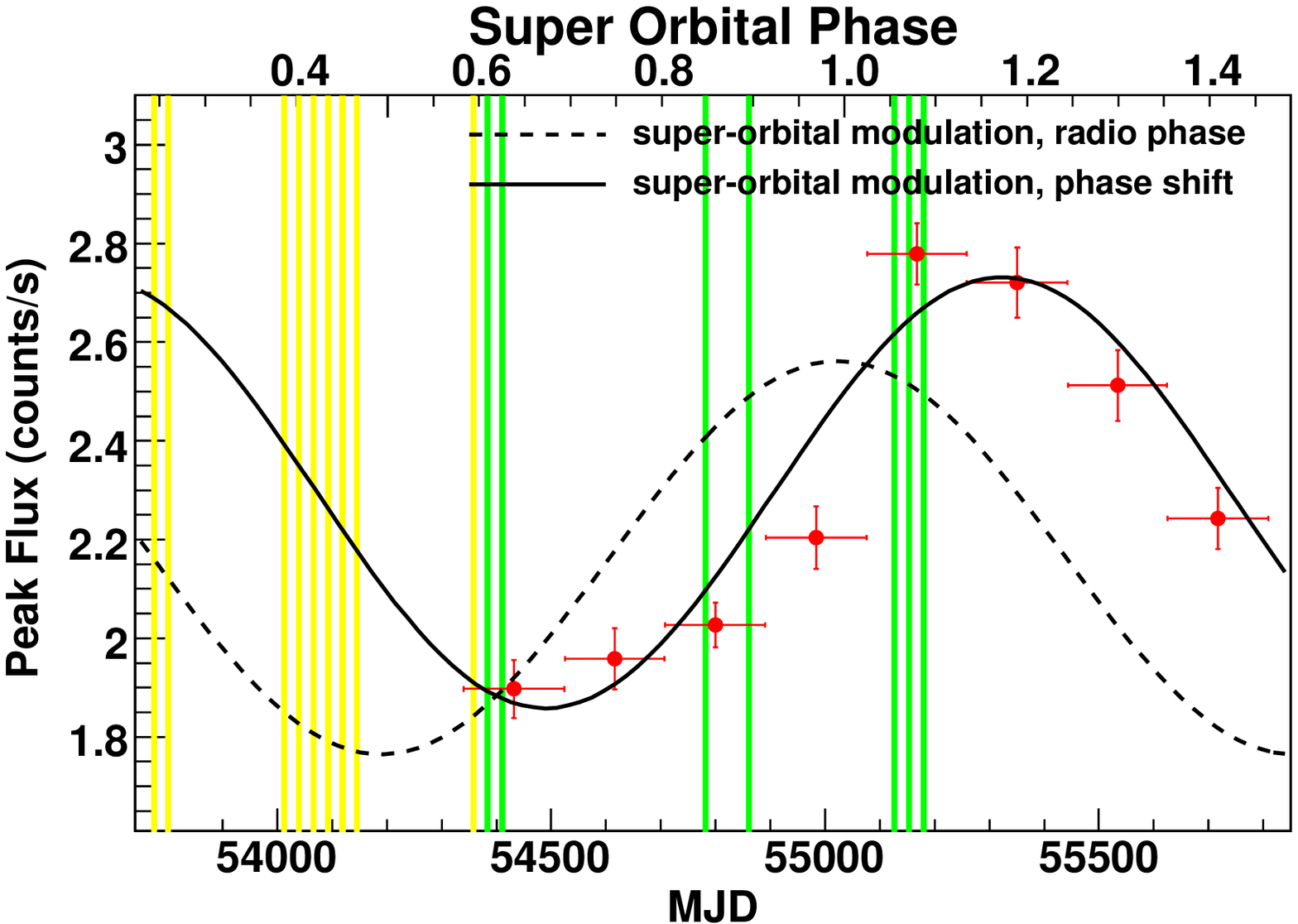}
  \includegraphics[angle=0, scale=0.4] {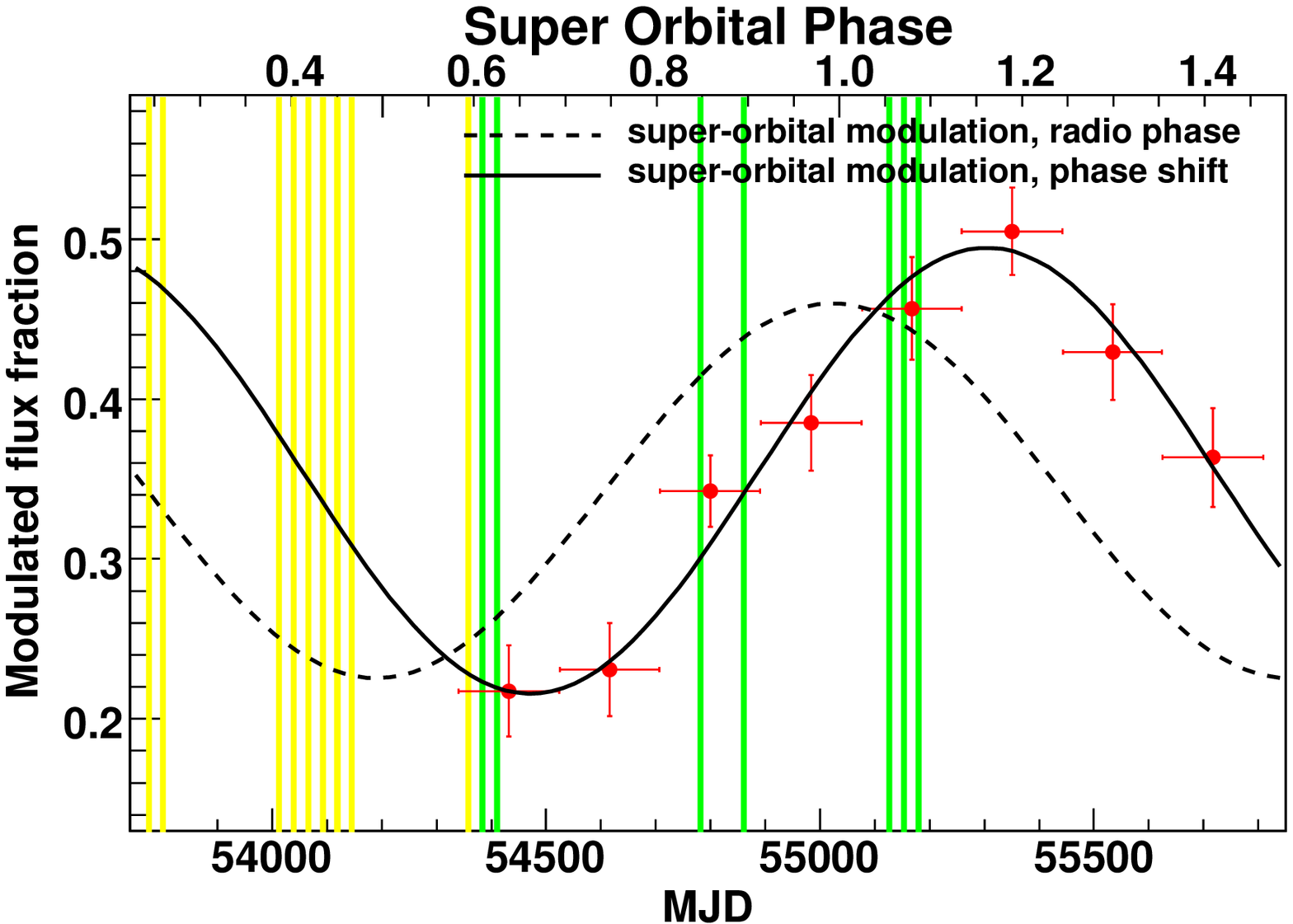}
 \caption{Left: Peak count rate  of the X-ray emission from \lsi\ as a function of time and the super-orbital phase.
 Right: modulated fraction, see text for details. The dotted line shows the sine fitting to the modulated flux fraction and peak flux with a period and phase fixed at the radio parameters (from Gregory 2002).
The solid curve stands for sinusoidal fit obtained by fixing the period at the 1667 days value, but letting the phase vary. The time bin corresponds to six months. The colored boxes represent the times of the TeV observations that covered the broadly-defined apastron region. The boxes in green denote the times when TeV observations are in low state while boxes in yellow are TeV observations in high state.
}
\label{super-orb}
\end{figure*}

Given a 6-months time bin (or approximately 6.8 orbits of the system), we take the peak X-ray flux in that period and compute the modulated flux fraction. The latter is defined as ($c_{max} - c_{min} )/(c_{max} + c_{min}$ ),
where $c_{max}$ and $c_{min}$ are the maximum and minimum count rates in the 3--30 keV orbital profile of that period,
after background subtraction. The minimum count rates are roughly constant around 1 count s$^{-1}$ all along the observation time. Results are shown in Figure \ref{super-orb}.
Table 1 presents the values of the reduced $\chi^2$ for fitting different models to the modulation fraction and the peak flux in X-rays.
It compares the results of fitting
a horizontal line, a linear fit, and two sinusoidal functions.
One of the latter has the same period and phase of the radio modulation (from Gregory 2002, labelled as `Radio' in Table 1,
dotted line in Figure 1). The other sine function has the same period as in radio but allowing for a phase shift from it (a solid line in Figure 1, labelled as `Shifted' in Table 1).
It is clear that there is variability in the data (and thus that a constant fit is unacceptable),
as well as that the sinusoidal description with a phase-shift is better than the linear one, which is obvious by visual inspection of Figure \ref{super-orb}. The phase shift derived by fitting the modulated fraction is  281.8 $\pm$ 44.6 days, corresponding in phase to $\sim 0.2$ of the 1667$\pm$8 days super-orbital period. The phase shift derived by fitting instead the maximum flux is 300.1 $\pm$ 39.1 days, and results compatible with the former.

\begin{table}[t]
\scriptsize
\begin{center}
\label{tab1}
\caption{Reduced $\chi^2$ for fitting different models to the modulation fraction and the peak flux in X-rays.
}
\vspace{5pt}
\small
\begin{tabular}{lllll}
\hline \hline
 & Constant & Linear & Radio & Shifted\\
\hline
Modulation Fraction & 88.2 / 7   & 38.0 / 6   &  42.1 / 6 & 1.1 / 5  \\
Peak Flux & 212.8 / 7   &  114.8 / 6  &  91.8 / 6 & 4.9 / 5  \\
\hline
\end{tabular}
\end{center}
\end{table}

\begin{figure*}[t]
\centering
  \includegraphics[angle=0, scale=0.35] {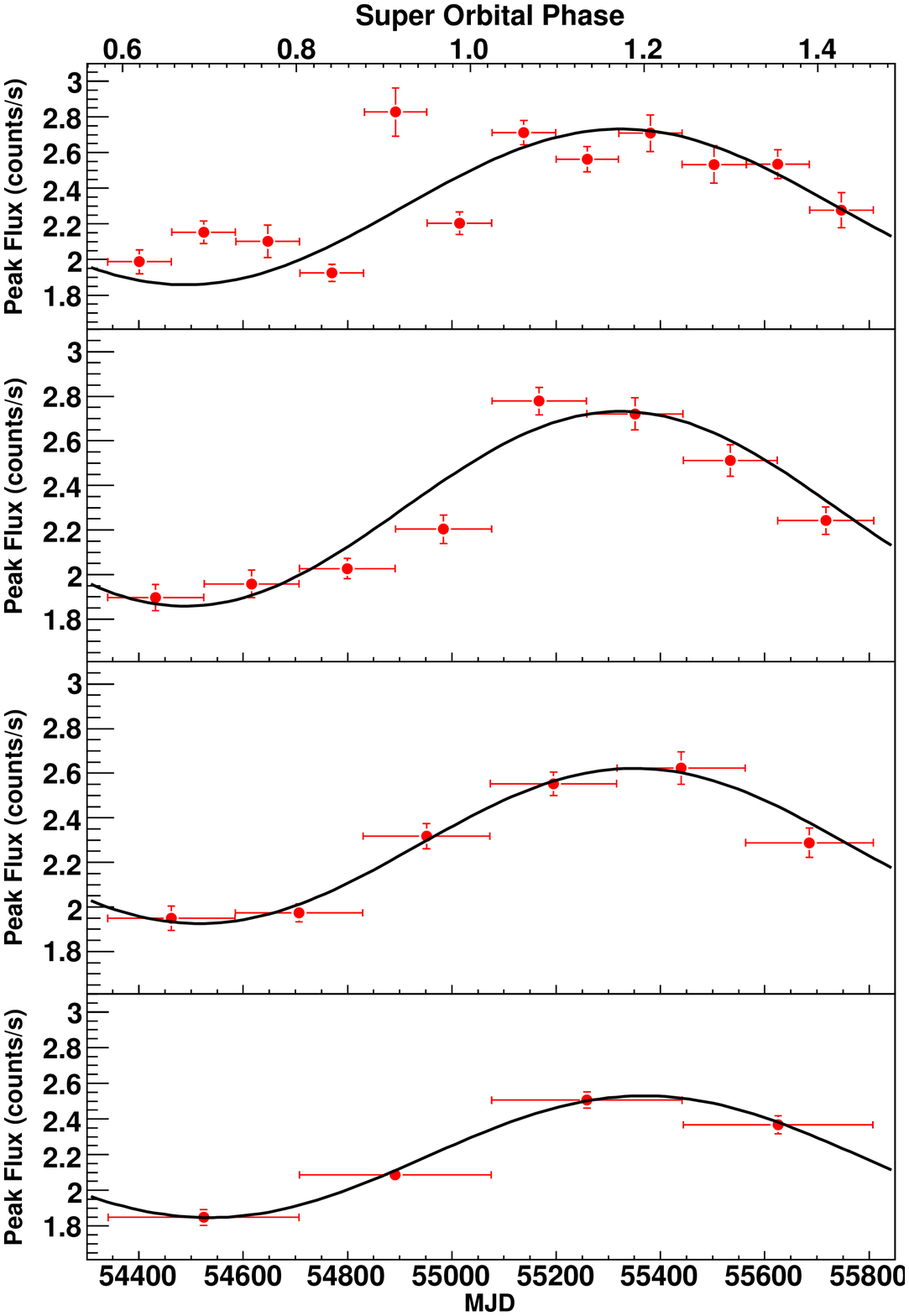}
  \includegraphics[angle=0, scale=0.35] {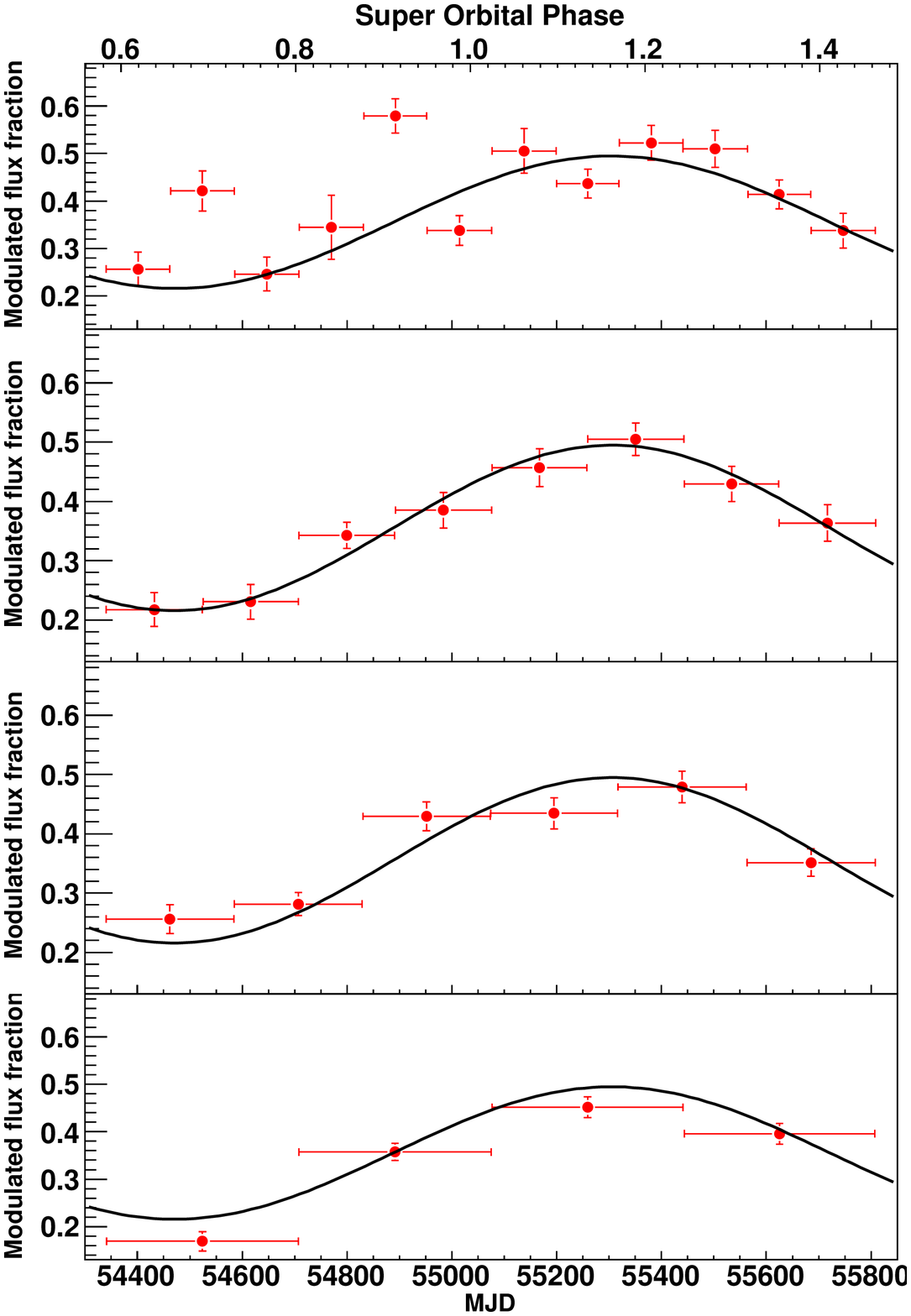}
 \caption{Peak flux (left) and modulation fraction (right) at different time scales.  From top to bottom we plot the results obtained by binning the data in 4, 6, 8, and 12 months bins. The black line shows the sine fitting with a 1667 days period fixed, as in the corresponding panels of Figure 1}
\label{scales}
\end{figure*}

Figure \ref{scales} analyzes whether there is anything special concerning the 6 months time bin just chosen to present the former results. It presents the peak flux and modulated fraction in different time bins, from 4 to 12 months. The black lines in Figure \ref{scales} shows the sine fitting with a 1667 days period fixed. The black lines in all left (right) panels are the same as the one depicted in the left (right) panel of Figure \ref{super-orb}.
It is interesting to note that the smaller the time bin the larger it is the scattering
around the sinusoidal fit, which can be understood as an effect of the increasing
similarity between the time bin itself and the orbital period of the system (of 26.4960$\pm$0.0028, Gregory 2002). Indeed, orbit-to-orbit variability is known to exist in our data, and can be similar or larger than the super-orbital induced variability at times (e.g.,
see the variations found for the same phase-bin in contiguous orbits in Figure 4 of Torres et al. 2010).
Thus, the shorter the time bin, the less likely it is that the super-orbital induced variation could be detected, which may be sub-dominant to the local-in-time changes. On the contrary, as soon as the integrated time bin is large enough in comparison with the orbital period of \lsi, the super-orbital variability is consistently observable.

This same fact makes  a direct comparison of the sinusoidal trend of the super-orbital X-ray modulation of Figure \ref{super-orb} with the
flux obtained from \lsi\
under short and isolated observations
untrustable. For example
the observations at soft X-rays conducted by  {\em XMM-Newton} (Neronov
\& Chernyakova 2006; Chernyakova et al. 2006, Sidoli et al. 2006), {\em Chandra}
(Paredes et al. 2007, Rea et al. 2010), {\em ASCA} (Leahy et al. 1997), {\em ROSAT}
(Goldoni \& Mereghetti 1995; Taylor et al. 1996), and {\em Einstein} (Bignami et al.
1981) were all too short to cover even a single full orbit. Similarly, earlier campaigns with PCA
during the month of March 1996 (Harrison et al.
2000,  Greiner and Rau 2001) are also too short.
Comparing their obtained flux values with the super-orbital trend would be meaningless since there is not enough time coverage to average out the possible orbit-to-orbit variations.

Finally, we note that we have also analyzed 15 years of \rxte-ASM data on \lsi, but the larger error bars
on the count rates preclude to obtain any conclusion from that dataset.

\section{Discussion}

  We notice that the X-ray long-term monitoring (2007--2011) on \lsi\  started about 7 years later from the end
of the campaign used to determine the super-orbital period in radio (1977--2000, see Gregory 2002).
We will assume then that the radio-determined super-orbital modulation of the source is, although possibly variable, active today with similar features as the ones claimed a decade earlier. This appears possible given that recent reports of variation in the orbital radio maxima are of only $\sim$0.1 in phase (see Trushkin \& Nizhelskij 2010).
Under such assumptions, we showed that there is a phase shift between the radio and the X-ray super orbital modulation.

Interestingly,
this shift is the same as the one hinted at between the radio and the H$\alpha$ line (Zamanov et al. 1999, Zamanov \& Mart\'i 2000).
Indeed,
the optical observations that covered the period 1989--1999
were fitted with a period of $\sim$1584 days (Zamanov et al. 1999), a value reported prior to the work by Gregory (2002), where the super-orbital period was refined to 1667$\pm$8 days. To investigate further the long-term modulation of \lsi\ at optical wavelengths and how it compares with the current findings in X-rays, we took the H$\alpha$ data from Table 4 of Paredes et al. (1994), Table 1 of Zamanov et al. (1999),  and Figure 1 of Zamanov et al. (2000), and as stated in Zamanov et al. (1999), considered an
error of 10\% for all the equivalent widths.
We performed an analysis similar to the one done for X-rays in the previous section, and derived and optical phase lag of 290.1$\pm$16.7 days with respect to the radio phase. Thus, the optical phase lag is coincident with the one derived at X-rays, although the observations at the two bands are about 8 years apart.
The stellar disk of Be stars are well known to grow larger as the equivalent width of the H$\alpha$ emission line increases
(e.g., Hanushik et al. 1988;  Grundstrom \& Gies 2006).
The optical variability has been most likely attributed to the cyclic variation of Be circumstellar disk. Thus, the possible coincidence with the X-ray phase lag suggests that the stellar disk may play an important role also for the X-ray emission, and probably for the higher-energy non-thermal emission of \lsi\ too.

The coincidence between the X-ray and optical shift with respect to the 1667 days radio modulation has to be taken with the necessary prudence prompted by it being based on non-simultaneous observations. In particular, it seems that the optical observations present the largest degree
of variation in time.
We checked that in addition of the H$\alpha$ measurements mentioned above, there are more recent ones
in the works of e.g., Liu et al. (2005), Grundstrom et al. (2007), Zamanov et al. (2007), and McSwain et
al. (2010). However,  the latter span 0.51 (at best, being usually much shorter) of
the super-orbital period, and as such we can not directly use them for a
comparison in long-terms. Nevertheless, they seem to hint at that the H$\alpha$ variability is not strictly periodic or at least at a changing amplitude.

Based (among other reasons discussed in Torres et al. 2011) on the analysis of a
the \swift-BAT detection of a short, magnetar-like burst from the direction of \lsi, we have proposed that the system's compact object is  a high magnetic field, slow period pulsar.
In that case, we proved that
the \lsi-system would most likely be subject to a flip-flop behavior, from a rotationally powered regime (in apastron, also known as ejector), to a propeller regime (in periastron) along each of the system's eccentric orbits.  The multi-wavelength phenomenology can be put in the context of the former model, and in particular, also the highest energy TeV emission, which has also shown low and high states which are apparently modulated by the same super-orbital period as well.
Within this model, we notice that
an increase in the accreted mass onto the compact object (unavoidably linked to the mass-loss rate of the star)
by a factor of a few\footnote{Estimations of the cyclical variations in the mass loss-rate from the Be star in \lsi\ are given as the ratio between maximal and minimal values obtained either from radio emission (a factor of 4 was determined by Gregory \& Neish 2002) or from H$\alpha$ measurements, which span from a factor of 5.6 (Gregory et al. 1989) to 1.5 (Zamanov et al. 1999).} can put the system in a permanent propeller stage along the orbit, including at the apastron region. This change of behavior for such an small change in mass loss rate can be the reason behind the evolution of the modulated fraction.
Indeed, using the formulae in Torres et al. (2011), and considering to simplify that the condition $R_m = R_{lc}$, where $R_m$ stands for the magnetic radius and $R_{lc}$ for the light cylinder,
establishes both the out-of-ejector and into-ejector condition,
one sees that the period--mass-loss--magnetic field
relation for the apastron of \lsi\ is
$
 \left( {P}/{ \rm 1\; s} \right) \sim a \! \times \!
   15 ( { B}/ {10^{14} \; {\rm G} } )^{4/7}  ( {\dot M_* }/ {10^{18} {\rm \; g \; s^{-1}} } )^{-2/7}
$
where $a$ represents a constant of order 1, and we have assumed an eccentricity of 0.6 and a semi-major axis of $6\times10^{12}$~cm. For periods shorter than the former, the system is in an ejector phase. For larger periods, it is in a propeller stage (see Torres et al. 2011 for details). High values of magnetic field and slow periods would make the transition possible: a cyclical change by a factor of a few in $\dot M_*$
can make the system to abandon the ejector phase  in apastron. For instance, under a variation by a
factor of 4 in $\dot M_*$,   a case that leads to a super-orbital induced transition is given by a magnetic field of $5 \times 10^{13}$ G, and a typical period of magnetars ($\sim 7$ s).
This may also happen for smaller values of the magnetic field but only in the case of a relatively long period. For instance, again under a variation by a factor of 4  in $\dot M_*$ and for $B=10^{12}$ G, the period should
be between 700 ms and 1s in order for the system to flip-flop in the super-orbital evolution, although no known pulsar in these parameter ranges
has a rotational energy in excess of $ 10^{36}$ erg s$^{-1}$ (ATNF Catalogue version: 1.43), which would be needed to account for
the multi-wavelength output of the system. Note in particular that the behavior of the \lsi\ system containing a pulsar with $B \sim 10^{12}$ G and $P < 700 $ ms would be unaffected by the cyclical variation of the mass-loss rate: it would act as an ejector in apastron along the whole super-orbital period.

The flip-flop mechanism can then be used to qualitatively explain why \lsi\ has entered in a low TeV state (see, e.g., Acciari et al. 2010) when at the maximum of the radio super-orbital variability, but perhaps also to explain why the modulated X-ray flux fraction
varies as we found in Figure \ref{super-orb}.
When the mass-loss rate is low,
the inter-wind shock formed at the collision region between the pulsar and the stellar wind would be present at the broad apastron region (and so will the TeV emission there), disappearing at periastron. In this situation, there are two contributors to the X-ray emission  along the orbit, expected to be roughly at the same level (e.g., Zamanov et al. 1999); the shock at apastron and the propeller at periastron, and the modulated fraction is consistently low.
When at the maximum of the mass-loss rate, the inter-wind shock may not form, and abundant TeV particles would not be produced since shocks at the magnetic radius are unable to reach TeV energies. Thus there is only one process generating the X-ray emission along the system's orbit, the propeller, and the modulated fraction is then maximum.
The exact position of the X-ray maximum along each of the orbits would depend on the local-in-time conditions
of the accreted mass onto the compact object, which established the relative weight of the two X-ray contributors,
and it is thus expected to vary beyond the super-orbital trend in short timescales, and not always be located at periastron.
However, given that the H$\alpha$ cycle represents the cyclical modulation of the mass loss rate, it would be natural to expect that the X-ray emission be correlated with it in long timescales (i.e., with how much mass is falling towards the compact object, e.g., see Bednarek 2009 or Bednarek \& Pabich 2011).

Zamanov et al. (2001) already discussed when the radio emission is expected to peak in each of the system orbits: The switch on of the ejector phase will activate the pulsar wind, creating a cavern around the neutron star which will start to expand. This means that the radio outburst will peak with some delay after the change of regimes, which is supposed to happen somewhere after the periastron, when the accretion rate onto the compact object diminishes enough.
In a cyclical variability of the mass loss rate of the star, the ejector-propeller transition moves in phase: at lower mass loss rates, the ejector will switch on earlier, and the radio outburst will peak at earlier orbital phases than at higher mass loss rates.  A generic TeV and radio anti-correlation is thus expected since the more mass fuels the propeller phase the more violent the radio outburst will be, and the less effectively the inter-wind shock  will generate TeV particles.

\begin{figure*}[t]
\centering
  \includegraphics[angle=0, scale=0.7]{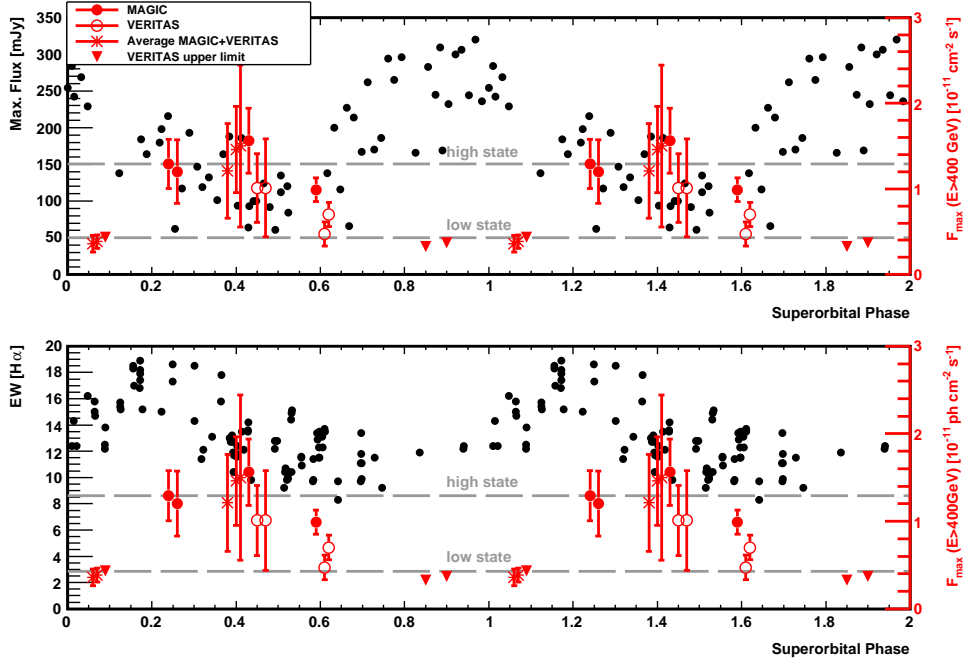}
 \caption{Peak flux per orbit in TeV shown in red (all of them happening in the 0.6--1.0 orbital phase range)
 as a function of superorbital phase, together with radio (top panel) and H$\alpha$ (bottom panel) data(black) as described in the text. The upper gray dashed line
 stands for the TeV flux level at discovery of the source in 2006, whereas the lower dashed line stands for 1/3 of this flux value. Two super-orbital phases are shown for clarity. Whenever there are both MAGIC and VERITAS compatible observations for the same orbit, they are averaged.
 }
\label{scales2}
\end{figure*}

The colored boxes in Figure \ref{super-orb} represent the times of the
TeV observations that covered the broadly-defined apastron region (from Albert et al. 2006, 2008, Anderhub et al. 2009, Aleksic et al. 2011; Acciari et al. 2008, 2009, 2010).
Those boxes colored in green denote the times for which the observations led only to imposing  an upper-limit or to a detection
with a flux that is about 3 times less than the one obtained at the discovery observations of 2006 (Albert et al. 2006), which defines the low state. The yellow boxes stand for those observations for which the level of the TeV emission was roughly compatible with the original discovery. There is a trend for finding a low TeV state towards the maximum of the super-orbital low-frequency cycles.
This is perhaps more clearly seen in Figure \ref{scales2}, where we plot the peak flux per orbit in TeV (all of them happening in the 0.6--1.0 orbital phase range)
 as a function of superorbital phase, together with radio and H$\alpha$ data.
However, the scarcity (and non-simultaneity) of the TeV coverage precludes reaching a definite conclusion on whether there is an anti-correlation of the TeV emission with the radio or with the H$\alpha$ curves. It would seem, however, that the TeV emission is rather anti-correlated with the radio flux and not with H$\alpha$, but this could not be quantitatively proven with the data at hand, especially given the caveats of dealing with non-contemporaneous observations.
A simultaneous optical-TeV campaign is needed to establish the nature of the anti-correlation. The latter would be  particularly useful for the forthcoming extrapolated radio maximum around October-November 2012.

\acknowledgements

We acknowledge support from the grants AYA2009-07391 and SGR2009-811, as
well as the Formosa Program TW2010005, by the National Natural Science Foundation of China via NSFC-10325313, 10521001, 10733010,11073021, and 10821061, the CAS key Project KJCX2-YW-T03,
and 973 program 2009CB824800. YPC thanks the Natural Science Foundation of China for support via NSFC-11103020 and 11133002. NR is supported by a Ramon y Cajal Fellowship.
We also acknowledge the use of the High Energy Astrophysics Science Archive Research Center (HEASARC), provided by NASA's Goddard Space Flight Center. JL acknowledges the hospitality of IEEC-CSIC, where this research was conducted.\\

\end{document}